\documentclass{article}
\usepackage{graphicx} 
\usepackage{amsthm}
\usepackage{amsmath}
\usepackage{amsfonts}
\usepackage{amssymb}
\usepackage{algorithm}
\usepackage{algorithmic}
\usepackage{booktabs}
\newtheorem{definition}{Definition}
\usepackage{float}
\usepackage{graphicx}
\newtheorem{lemma}{Lemma}
\newtheorem{theorem}{Theorem}

\title{Fast \(\ell_\infty\)-Reduction of Rank-2 Lattices in $\mathbb{R}^n$ via a New Reduced-Basis Notion}
\author{Bo Pang, Shaojing Zhang, Chengliang Tian \thanks{Corresponding author: \texttt{tianchengliang@qdu.edu.cn}}}

\begin{document}
\maketitle

\begin{abstract}
The shortest vector problem (SVP) under the infinity norm 
($\ell_\infty$-norm) is an important lattice problem with applications 
in lattice-based cryptography. In particular, the security analysis of 
the post-quantum signature standard ML-DSA highlights the importance of 
finding short solutions satisfying infinity-norm constraints, rather 
than only Euclidean norm bounds. However, most existing lattice 
reduction algorithms are developed for the Euclidean norm, while 
dedicated studies under the infinity norm remain limited.

In this work, we focus on rank-two lattices in $\mathbb{R}^n$ and 
introduce a new notion of reduced basis under the infinity norm. Based 
on this definition, we propose an efficient lattice basis reduction 
algorithm that performs a multi-dimensional analogue of Euclidean 
division using a dominant coordinate of the basis vectors. Each 
non-terminal iteration guarantees a factor-two decrease of the maximum 
infinity norm of the basis vectors. We prove that the algorithm 
terminates and achieves a bit complexity of $O(nT^2)$, matching the 
asymptotic complexity order of the classical Lagrange--Gauss reduction. 
Moreover, the proposed algorithm saves at least $2n$ multiplications per iteration compared with the classical approach. 
Experiments demonstrate significant runtime improvements, especially 
for high-dimensional lattices. This work provides a foundation for 
developing lattice reduction techniques under the infinity norm.
\end{abstract}

\section{Introduction}
\label{sec:introduction}

Lattice basis reduction is a fundamental algorithmic tool in computational number theory, integer optimization, and cryptography. It is used in the analysis of classical public-key cryptosystems and in the security analysis of lattice-based schemes, where the ability to find short integer vectors is closely related to the difficulty of underlying lattice problems \cite{Ngu01,micciancio2002complexity}. The geometry induced by the norm is an essential part of this algorithmic picture: a vector that is short in one norm need not be shortest in another, and the structure of a reduction step can change substantially with the chosen norm. Although Euclidean reduction has been studied extensively, coordinate-wise constraints arising in applications also motivate the direct study of non-Euclidean norms, in particular the infinity norm $\|\mathbf{x}\|_\infty=\max_i |x_i|$. For example, coordinate bounds expressed through the infinity norm occur in the signing procedure and security analysis of lattice-based signature schemes such as ML-DSA (formerly CRYSTALS-Dilithium) \cite{bai2021crystals,wang2025heuristic}. The security problem for ML-DSA is structured and high-dimensional, rather than a rank-two SVP instance; nevertheless, such applications illustrate why algorithms tailored to the infinity norm are of independent interest.

This work studies basis reduction for rank-two lattices under the infinity norm. A rank-two lattice can be represented by two linearly independent vectors $\mathbf{a},\mathbf{b}\in\mathbb{Z}^{n}$, with basis matrix $\mathbf{B}=[\mathbf{a}\ \mathbf{b}]\in\mathbb{Z}^{n\times 2}$. The planar full-rank case corresponds to $n=2$, whereas for $n>2$ the same rank-two lattice is embedded in a higher-dimensional ambient space. This distinction is important: algorithms designed around the two coordinates of a planar basis do not automatically generalize to arbitrary coordinate dimension, while a method for rank-two lattices in $\mathbb{R}^n$ must account for all $n$ coordinates. Our goal is to develop a simple, exact reduction rule for this latter setting, with a shortest-vector guarantee and without inner-product computations.

\paragraph{Classical Euclidean reduction and the choice of the reduction coefficient.}
For a full-rank planar lattice with basis $[\mathbf{a}\ \mathbf{b}]\in\mathbb{Z}^{2\times 2}$, the classical Lagrange--Gauss algorithm solves the shortest vector problem (SVP) exactly \cite{lagrange1773recherches,gauss1801disquisitiones}. Suppose that $\|\mathbf{a}\|_2\geq\|\mathbf{b}\|_2$. A basic step replaces $\mathbf{a}$ by $\mathbf{c}=\mathbf{a}-q\mathbf{b}$ and then exchanges the two basis vectors. The coefficient is chosen to minimize $\|\mathbf{a}-q\mathbf{b}\|_2$ over integers $q$. Expanding the squared norm shows that its real-valued minimizer is $\langle\mathbf{a},\mathbf{b}\rangle/\langle\mathbf{b},\mathbf{b}\rangle$, and consequently an optimal integer coefficient is
\begin{equation}
  q=\left\lfloor\frac{\langle\mathbf{a},\mathbf{b}\rangle}
  {\langle\mathbf{b},\mathbf{b}\rangle}\right\rceil,
  \label{eq:euclidean-q}
\end{equation}
where $\lfloor\cdot\rceil$ denotes rounding to a nearest integer, with a fixed tie-breaking rule. Thus, for the Euclidean norm, choosing $q$ reduces to one inner-product ratio followed by integer rounding. Repeated reduction yields a Lagrange-Gauss-reduced basis whose shorter vector is a shortest nonzero vector of the lattice.

The analogy between Lagrange-Gauss reduction and the integer Euclidean algorithm has led to a long line of work on iteration bounds and bit complexity. Lagarias analyzed the complexity of related reduction procedures for integral binary quadratic forms \cite{Laga80}; Vall\'ee derived sharp bounds on the number of iterations of Gauss reduction \cite{Val91}; and Nguyen and Stehl\'e proved that the greedy Euclidean reduction algorithm has quadratic bit complexity, without fast integer multiplication, in dimensions at most four \cite{NS09}. A separate line of research sought to simplify the arithmetic performed in each reduction step. In particular, Pornin proposed a practical variant in which the reduction multiplier is constrained to be a power of two, so that multiplication by the multiplier can be implemented by shifts. This reduces the cost of an individual step at the price of potentially performing more steps \cite{Por20}. Such variants illustrate that the best coefficient for minimizing the next vector norm need not always be the cheapest coefficient to compute or apply.

\paragraph{From individual quotients to recursive blocks of reductions.}
A more substantial acceleration comes from computing several reduction steps together rather than applying the Euclidean-style quotient operation one step at a time. Yap studied fast unimodular reduction for planar integer lattices and developed a Half-Gaussian method inspired by the half-GCD algorithm \cite{yap1992fast}. In addition to the centered nearest-integer rule in Eq.~\eqref{eq:euclidean-q}, reduction algorithms can use a floor quotient, such as $q=\lfloor\langle\mathbf{a},\mathbf{b}\rangle/\langle\mathbf{b},\mathbf{b}\rangle\rfloor$, to maintain an uncentered reduction convention and an acute-angle invariant. Yap's CRS and Half-Gaussian approaches exploit the relation between sequences of such steps and products of unimodular transformation matrices. Independently, Sch\"onhage developed fast reduction and composition methods for binary quadratic forms. His partial-reduction strategy uses a thresholded quotient of the form
\begin{equation}
  q=\left\lfloor\frac{\langle\mathbf{a},\mathbf{b}\rangle-s}
  {\langle\mathbf{b},\mathbf{b}\rangle}\right\rfloor,
  \qquad s=2^m>0,
  \label{eq:thresholded-q}
\end{equation}
which permits a block of progress to be inferred from leading bits rather than by carrying out every full-precision reduction step \cite{Sch91}. These recursive approaches achieve $O(M(T)\log T)$-type bit-complexity bounds for fast planar reduction, where $T$ is the input bit length and $M(T)$ is the bit complexity of multiplying two $T$-bit integers. Ryan and Heninger subsequently provided a detailed implementation of fast reduction techniques for binary quadratic forms \cite{RH23}. The common principle is to accumulate a block of elementary operations in a small unimodular matrix and apply that matrix to the full-precision input, much as in integer HGCD.

\paragraph{The infinity norm and planar special cases.}
Under the infinity norm, the Euclidean projection formula in Eq.~\eqref{eq:euclidean-q} is no longer available. The generic coefficient-selection problem becomes
\begin{equation}
  q^*\in\arg\min_{q\in\mathbb{Z}}
  \|\mathbf{b}-q\mathbf{a}\|_\infty
  =\arg\min_{q\in\mathbb{Z}}\max_{1\leq i\leq d}|b_i-qa_i|.
  \label{eq:linfty-q}
\end{equation}
The evolution from the Euclidean case to the infinity norm therefore involves not merely replacing one norm by another, but also replacing a closed-form projection quotient with a different optimization problem.

For planar integral lattices, Eisenbrand showed that an infinity-norm shortest vector can be found using a constant number of extended-GCD computations, one common-convergent computation, and a constant number of additional arithmetic operations, thereby connecting the planar $\ell_\infty$ SVP to continued fractions and fast GCD computation \cite{EIS01}. Subsequent work on the minimal rational fraction representation of binary sequences exploited a special HNF representation and developed an efficient reduction procedure with an explicit termination criterion \cite{CTJ22}. In the planar case, the coordinate sign pattern makes it possible to write the continuous minimizer explicitly. When $a_1a_2\neq 0$, let $r_i=b_i/a_i$ and $w_i=|a_i|$. Then
\begin{equation}
 t^*=\frac{w_1r_1+w_2r_2}{w_1+w_2}
 =\frac{\operatorname{sgn}(a_1)b_1+\operatorname{sgn}(a_2)b_2}
 {|a_1|+|a_2|},
 \label{eq:planar-linfty-continuous}
\end{equation}
which is the weighted minimax point between the two ratios $r_1$ and $r_2$. Depending on their signs, the numerator becomes a sum or a signed difference of the coordinate magnitudes. An optimal integer coefficient can then be selected by comparing $\lfloor t^*\rfloor$ and $\lceil t^*\rceil$. Degenerate coordinate cases require separate handling. This coordinate-specific formula avoids inner products, but it relies on the two-dimensional sign and magnitude structure. Wu and Xu explored another special family of planar lattices arising from Qin Jiushao's algorithm; their method uses state matrices and unimodular transformations to extract a shortest vector from a structured recurrence \cite{wu2023qin}. These results are efficient for their respective planar or structured settings, but their formulas do not by themselves furnish a coefficient-selection rule for arbitrary vectors in $\mathbb{Z}^n$.

\paragraph{Generalized Gauss reduction for rank-two lattices in arbitrary norms.}
A general framework that goes beyond planar coordinates was established by Kaib and Schnorr, who generalized Gauss reduction from the Euclidean norm to an arbitrary norm on $\mathbb{R}^n$ \cite{Kaib1996}. For a rank-two lattice $L=\mathbb{Z}\mathbf{a}+\mathbb{Z}\mathbf{b}$, they call a basis reduced when
\begin{equation}
  \|\mathbf{a}\|,\|\mathbf{b}\|
  \leq \|\mathbf{a}-\mathbf{b}\|
  \leq \|\mathbf{a}+\mathbf{b}\|.
  \label{eq:generalized-gauss-reduced}
\end{equation}
They prove that the lengths of the vectors in such a reduced basis attain the first two successive minima under the chosen norm. Their generalized Gauss algorithm repeatedly chooses an integer coefficient minimizing the norm of a vector of the form $\mathbf{b}-q\mathbf{a}$, followed by suitable sign changes and basis exchanges. For a general norm, this minimization can be treated using a norm oracle and a search procedure; its implementation depends on the structure of the norm.

The computation of $q$ is explicit for several important norms. For $\ell_2$, it is given by the nearest-integer projection formula in Eq.~\eqref{eq:euclidean-q}. For $\ell_1$, the real-valued minimizer of $\sum_i|b_i-ta_i|$ is a weighted median of the ratios $b_i/a_i$, with weights $|a_i|$. For $\ell_\infty$, the function
\[
  f(t)=\max_{1\leq i\leq d}|b_i-ta_i|
\]
is the upper envelope of the $2n$ lines $b_i-a_it$ and $-b_i+a_it$. Sorting the lines by slope and scanning their upper envelope finds a real minimizer in $O(n\log n)$ arithmetic operations; comparison of the two adjacent integers around a minimizer yields an optimal integer coefficient. This gives a generic way to implement a reduction step in arbitrary ambient dimension, but it also shows that coefficient selection under $\ell_\infty$ has a different computational structure from the single inner-product ratio used under $\ell_2$.

Kaib's fast variant addresses the bit complexity of the generalized Gaussian algorithm for inputs $\mathbf{a},\mathbf{b}\in\mathbb{Z}^n$ under the $\ell_1$, $\ell_2$, and $\ell_\infty$ norms \cite{Kaib1994}. The algorithm uses high-bit approximations, recursive calls that compute blocks of reduction steps, unimodular transformation matrices, and controlled restoration of discarded low bits. For coordinate bit length at most $T$, it achieves $O(nM(T)\log T)$ bit operations under $\ell_\infty$ and $O(n\log n\,M(T)\log T)$ under $\ell_1$ and $\ell_2$. Thus, arbitrary-dimensional rank-two reduction under the infinity norm is not an unexplored problem: Kaib's work already provides a fast general-norm framework and a corresponding bit-complexity analysis. The question pursued here is instead whether a different coordinate-wise reduction criterion can avoid the generic norm-minimizing coefficient search and yield a simpler iteration based on elementary integer operations, while preserving an exact shortest-vector guarantee.

\paragraph{Recent planar algorithms and the remaining motivation.}
Zhao et al. recently revisited the SVP for full-rank planar lattices, introduced a new reduced-basis criterion, and proposed the adaptive CrossEuc algorithm together with HVec, a vectorized generalization of integer HGCD. By iteratively applying HVec, their HVecSBP algorithm obtains a reduced basis in $O(M(T)\log T)$ time for general planar input bases, without requiring a preliminary conversion to HNF \cite{zhao2025algorithms}. Their experiments report substantial improvements over the tested baselines, particularly for HNF inputs. These results demonstrate the continuing benefit of designing reduction criteria around the structure of the target norm and of accumulating multiple steps through transformation matrices. However, their setting remains the planar full-rank case, whereas the present work considers two vectors with an arbitrary number of coordinates.

The preceding literature establishes two complementary facts. First, classical Gauss reduction and its HGCD-inspired variants provide exact and highly efficient algorithms for planar Euclidean lattices, while specialized continued-fraction methods solve important planar $\ell_\infty$ instances. Second, Kaib and Schnorr's generalized theory, together with Kaib's fast variant, already covers rank-two lattices embedded in higher-dimensional spaces and includes exact shortest-vector guarantees under $\ell_\infty$. Consequently, our motivation is not to claim the first arbitrary-dimensional algorithm for this problem. Rather, we seek a complementary reduction rule that is directly adapted to the coordinate structure of the infinity norm and avoids inner-product computations and the generic per-step minimization of $\|\mathbf{b}-q\mathbf{a}\|_\infty$. This distinction is algorithmic: the goal is to obtain a simpler implementable procedure with a transparent descent argument and an exact SVP guarantee, rather than to replace the existing general-norm theory.

 The main contributions of this work are summarized as follows:
\begin{itemize}
  \item We introduce a new reduced-basis criterion for rank-two lattices in $\mathbb{Z}^n$ under the $\ell_\infty$ norm and prove that every basis satisfying the criterion contains a shortest nonzero lattice vector.
  \item We design a reduction algorithm based on coordinate-wise sign and magnitude relations. It avoids inner-product computations and uses Euclidean-division-like updates on selected coordinates. We prove correctness and show that every nonterminal iteration decreases the maximum infinity norm of the current basis vectors by at least a factor of two. Its bit complexity is $O(nT^2)$, where $T$ is the maximum bit length of the input coordinates.
  \item We implement the algorithm and evaluate it across different ambient dimensions and degrees of linear dependence between the input basis vectors. The experiments compare its practical performance with classical Lagrange--Gauss reduction and assess the effect of dimension and input geometry.
\end{itemize}

The remainder of this paper is organized as follows. 
Section 2 reviews preliminaries on lattice theory, SVP, and classical Lagrange reduction. 
Section 3 introduces the proposed reduction definition under the infinity norm and proves its theoretical properties. 
Section 4 presents the proposed reduction algorithm together with correctness and complexity analyses. 
Section 5 reports experimental results and performance comparisons. 
Finally, Section 6 concludes the paper.

\section{Preliminaries}
This section introduces the basic notations used in this paper and briefly reviews the fundamental concepts of lattices, the SVP and the classical reduction algorithm.

\subsection{Notations and Terminologies}
Throughout this paper, lowercase bold letters (e.g., $\mathbf{a}, \mathbf{b}$) are used to denote column vectors, and the $i$-th component of a vector $\mathbf{b}$ is denoted by $b_i$. Uppercase bold letters (e.g., $\mathbf{B}$) represent matrices, which are formed by column vectors serving as a lattice basis. The transpose of a matrix 
$\mathbf B$ is denoted by $\mathbf B^T$.

For any real number $x$, its sign function is denoted by
$$
\operatorname{sgn}(x)=
\begin{cases}
1,&x\geq0,\\
-1,&x<0.
\end{cases}
$$
In particular, we use $\lfloor x \rfloor_o$ to denote the ``rounding towards zero'' operation for a real number $x$, defined as follows:
$$
\lfloor x \rfloor_o = 
\begin{cases} 
\lfloor x \rfloor, & \text{if } x \ge 0 \\ 
\lceil x \rceil, & \text{otherwise} 
\end{cases}
$$

Unless explicitly stated otherwise, the vector length $\|\mathbf{x}\|$ referred to in this paper defaults to the infinity norm ($\ell_\infty$ norm), defined as the maximum absolute value of its components: $\|\mathbf{x}\| = \max_{1 \le i \le n} |x_i|$. 
Furthermore, for any $n$-dimensional vector $\mathbf{x}$, it satisfies the relation with the classical Euclidean norm ($\ell_2$ norm) bounded by $\|\mathbf{x}\|_\infty \le \|\mathbf{x}\|_2 \le \sqrt{n}\|\mathbf{x}\|_\infty$.

\subsection{Lattice and related properties}
\begin{definition}[\cite{micciancio2002complexity}]
Let $\mathbf{b}_1, \mathbf{b}_2, \dots, \mathbf{b}_m \in \mathbb{R}^n$ ($m \le n$) be a set of linearly independent vectors. A lattice of rank $m$ generated by them is defined as the set of all their integer linear combinations : 
$$ \mathcal{L} = L(\mathbf{b}_1, \dots, \mathbf{b}_m) = \left\{ \sum_{i=1}^m z_i \mathbf{b}_i \;\middle|\; z_i \in \mathbb{Z} \right\}. $$
Specially, when $m=n$, we say  $\mathcal{L}(\mathbf{B})$ is full-rank
\end{definition}

The matrix $\mathbf{B} = [\mathbf{b}_1, \dots, \mathbf{b}_m] \in \mathbb{R}^{n \times m}$ is called a \textbf{basis} of the lattice. The same lattice can be generated by different bases. If $\mathbf{B}'$ is another basis of the same lattice $\mathcal{L}$, there exists a unimodular matrix $\mathbf{U} \in \mathbb{Z}^{m \times m}$ \cite{lenstra1982factoring} (i.e., an integer matrix with $\det(\mathbf{U}) = \pm 1$) such that $\mathbf{B}' = \mathbf{B}\mathbf{U}$.

While the definitions above apply to arbitrary dimensions, the scope of this paper is primarily focused on rank-2 lattice embedded in $n$ dimensions (i.e., rank-2 lattices where $m=2$). In this context, the basis simply consists of a pair of vectors $\mathbf{B} = [\mathbf{b}_1, \mathbf{b}_2]$. 


\begin{definition}[\cite{micciancio2002complexity}](Successive Minima)
For a lattice $\mathcal{L}$ of rank $m$, the $i$-th ($1\le i \le m$) successive minimum
$\lambda_i(\mathcal{L})$ is defined as the smallest real number $r>0$
such that the ball centered at the origin with radius $r$
contains at least $i$ linearly independent lattice vectors.
\[
\lambda_i(\mathcal{L}) = \inf \{r > 0 : \dim(\operatorname{span}(\mathcal{B}(\mathbf{o},r) \cap \mathcal{L})) \geq i\},
\]where $\mathcal{B}(\mathbf{o},r) = \{\mathbf{x} \in \mathbb{R}^n : \|\mathbf{x}\| < r\}$ represents the open ball of radius $r$ centered at the origin.
\end{definition}

\begin{definition}[\cite{micciancio2002complexity}](Shortest Vector Problem, SVP)
Given a basis $\mathbf{B} \in \mathbb{Z}^{n \times m}$, find a nonzero lattice vector $\mathbf{Bx}$ (with $\mathbf{x} \in \mathbb{Z}^m \setminus \{\mathbf{0}\}$) such that $\|\mathbf{Bx}\| \le \|\mathbf{By}\|$ for any other $\mathbf{y} \in \mathbb{Z}^m \setminus \{\mathbf{0}\}$.
\end{definition}

It is well established that finding the shortest vector in high-dimensional lattices is computationally hard and is NP-hard to approximate within certain factors~\cite{micciancio2001shortest}.

\subsection{Lagrange-Gauss Reduced Basis and Algorithm}
The Lagrange algorithm (also known as Gauss reduction in some references) is one of the earliest exact reduction algorithms for rank-2 lattices in \(\mathbb{R}^2\) \cite{gauss1801disquisitiones,lagrange1773recherches}. Its iterative structure is closely related to the Euclidean algorithm, as it repeatedly shortens the basis vectors through subtraction and orthogonal projection. In their book, Micciancio and Goldwasser \cite{micciancio2002complexity} further generalized and established the mathematical criterion for reduced bases of rank-2 lattices in \(\mathbb{R}^n\) under any computable norm.

\begin{definition}[Lagrange–Gauss reduced basis \cite{lagrange1773recherches,micciancio2002complexity}]\label{def:Lag}
A basis \(\mathbf{B} = [\mathbf{a}, \mathbf{b}]\in\mathbb{R}^{n\times 2}\) of a lattice \(\mathcal{L}\) is called \emph{Lagrange--Gauss reduced}  if
\[
\max\{\|\mathbf{a}\|,\|\mathbf{b}\|\} \le \min\{\|\mathbf{a}+\mathbf{b}\|,\ \|\mathbf{a}-\mathbf{b}\|\}.
\]
\end{definition}

This criterion is universal: it applies to any computable norm $\|\cdot\|$, in particular the Euclidean norm and the infinity norm studied in this paper. Moreover, it implies the following well‑known result.

\begin{lemma}[\cite{micciancio2002complexity}]\label{lem:Lag}
If \(\mathbf{B}=[\mathbf{a},\mathbf{b}]\) is Lagrange--Gauss reduced, then for any computable norm \(\|\cdot\|\) we have
\[
\min\{\|\mathbf{a}\|,\|\mathbf{b}\|\} = \lambda_1(\mathcal{L})
\quad\text{and}\quad
\max\{\|\mathbf{a}\|,\|\mathbf{b}\|\} = \lambda_2(\mathcal{L}) .
\]
\end{lemma}
 
To transform any initial basis $[\mathbf{a}, \mathbf{b}]$ into a Lagrange-Gauss reduced basis, the processing flow can be described by the following pseudo-code, which is divided into an initial adjustment (\textit{start}) and a core reduction loop (\textit{loop}):

\begin{algorithm}[H]
\caption{Lagrange-Gauss Reduction Algorithm}
\begin{algorithmic}[1]\label{alg:LG}
\REQUIRE Two linearly independent vectors $\mathbf{a}$ and $\mathbf{b}$
\ENSURE A reduced basis of the lattice $\mathcal{L}([\mathbf{a}, \mathbf{b}])$

\textbf{(start):}
\STATE \textbf{if} $\|\mathbf{a}\| > \|\mathbf{b}\|$ \textbf{then} $\text{swap}(\mathbf{a}, \mathbf{b})$
\STATE \textbf{if} $\|\mathbf{a} - \mathbf{b}\| > \|\mathbf{a} + \mathbf{b}\|$ \textbf{then} $\mathbf{b} := -\mathbf{b}$
\STATE \textbf{if} $\|\mathbf{b}\| \le \|\mathbf{a} - \mathbf{b}\|$ \textbf{then return} $[\mathbf{a}, \mathbf{b}]$
\STATE \textbf{if} $\|\mathbf{a}\| \le \|\mathbf{a} - \mathbf{b}\|$ \textbf{then go to} (loop)
\STATE \textbf{if} $\|\mathbf{a}\| = \|\mathbf{b}\|$ \textbf{then return} $[\mathbf{a}, \mathbf{a} - \mathbf{b}]$
\STATE $[\mathbf{a}, \mathbf{b}] := [\mathbf{b} - \mathbf{a}, \mathbf{a}]$

\textbf{(loop):}
\STATE Find $q \in \mathbb{Z}$ such that $\|\mathbf{b} - q\mathbf{a}\|$ is minimal, $\mathbf{b} := \mathbf{b} - q\mathbf{a}$ \label{keystep}
\STATE \textbf{if} $\|\mathbf{a} - \mathbf{b}\| > \|\mathbf{a} + \mathbf{b}\|$ \textbf{then} $\mathbf{b} := -\mathbf{b}$
\STATE $\text{swap}(\mathbf{a}, \mathbf{b})$
\IF{$[\mathbf{a}, \mathbf{b}]$ is reduced}
    \RETURN $[\mathbf{a}, \mathbf{b}]$
\ELSE
    \STATE \textbf{go to} (loop)
\ENDIF
\end{algorithmic}
\end{algorithm}
Also, the above reduction is applicable to any computable norm. The key difference lies in Step \ref{keystep} within each loop, which is the most time‑consuming step. A general method for finding the optimal multiplier \(q\) is binary search; however, for the classical \(\ell_2\) norm, the optimal multiplier can be determined directly by the closed‑form expression  
\[
q = \left\lfloor \frac{\langle \mathbf{a}, \mathbf{b} \rangle}{\langle\mathbf{a},\mathbf{a}\rangle} \right\rceil,
\]  
where \(\lfloor \cdot \rceil\) denotes rounding to the nearest integer.


It is easy to deduce that, the output basis perfectly meets the definition of Lagrange reduction basis $\|\mathbf{a}\|\le\|\mathbf{b}\|\le\|\mathbf{a}-\mathbf{b}\|\le\|\mathbf{a}+\mathbf{b}\|$. However, when processing extremely large integer inputs, the classical algorithm frequently executes highly expensive high-precision arithmetic operations (to find the optimal integer $q$ in the loop), often encountering severe performance bottlenecks. This motivates the investigation of alternative reduction strategies that avoid expensive inner-product computations.

\section{New Reduced Basis}\label{sec:RedB}

In classical lattice theory, the Lagrange reduced basis is inherently tied to the Euclidean metric, which involves computing inner products and entails significant computational overhead for large bit-length integers. To achieve optimal efficiency, it is imperative to shift the reduction criteria to the infinity norm ($\ell_\infty$). However, simply replacing the norm in the classical definition does not yield an algorithmically friendly termination condition. Therefore, in this section, we propose a novel, coordinate-wise definition of a reduced basis explicitly tailored for the infinity norm. This new definition bypasses the need for inner products, relying solely on the signs and relative magnitudes of the vector components.
\begin{definition}\label{def:Eucredn}
A lattice basis \begin{equation}\label{eq:basis}
   \mathbf{B}=[\mathbf{a}\ \mathbf{b}]\in\mathbb{R}^{n\times 2}, 
\end{equation} with $\mathbf a=(a_1\ a_2\ \cdots\ a_n)^T$, $\mathbf b=(b_1\ b_2\ \cdots\ b_n)^T$, $\|\mathbf{a}\|=|a_i|$ and $\|\mathbf{b}\|=|b_j|$ 
is called reduced if one of the following conditions holds.
     \begin{align*}
	 &(\star_1): \ a_ib_i a_jb_j\le 0.
	 \\
  &(\star_2):\ a_ib_i a_jb_j > 0\wedge \exists k \neq i,j(a_ib_i a_kb_k\le 0\wedge |a_k|+|b_k|\ge\min\{\|\mathbf{a}\|,\|\mathbf b\|\}).  
	 \end{align*}
\end{definition}
The following theorem the relationship between the proposed reduction condition and the successive minima of the lattice $\mathcal{L}=L(\mathbf{B})$. 
\begin{theorem}
    Let $\mathcal{L}=L(\mathbf{B})$ be a lattice with a reduced basis $\mathbf{B}$ as in Definition \ref{def:Eucredn} and $\|\mathbf a\|\le \|\mathbf b\|$.Then $\|\mathbf{a}\| = \lambda_1(\mathcal{L})$, and one of the following holds for $\|\mathbf{b}\|$ :
    \begin{equation}
        \|\mathbf{b}\| = \lambda_2(\mathcal{L}),\ \text{or}\  \|\mathbf{a}\| \le \|\mathbf{a}+\mathbf{b}\| < \|\mathbf{b}\|,\ \text{or}\  \|\mathbf{a}\| \le \|\mathbf{a}-\mathbf{b}\| < \|\mathbf{b}\|.
    \end{equation}
\end{theorem}
\begin{proof} 
  \noindent\textbf{Argument for the length of $\mathbf{a}$.}  That is, for any non-zero lattice vector $\mathbf{v}=z_1\mathbf{a}+z_2\mathbf{b}=(z_1a_1+z_2b_1\cdots\ z_1a_i+z_2b_i\ \cdots\ z_1a_j+z_2b_j\ \cdots\ z_1a_n+z_2b_n)^T$, we will deduce 
\begin{align}
    \lambda_1(\mathcal{L})=\|\mathbf{a}\|\le \|\mathbf{v}\|.
\end{align}
Clearly, if $z_2=0$, then $\|\mathbf{v}\|\ge|z_1|\,\|\mathbf{a}\|\ge\|\mathbf{a}\|$, and,
if $z_1=0$, then $\|\mathbf{v}\|\ge|z_2|\,\|\mathbf{b}\|\ge\|\mathbf{b}\|\ge\|\mathbf{a}\|$.

Now suppose $z_1z_2\neq 0$.
We distinguish two main situations according to the sign of $z_1z_2$ and the defining conditions of a reduced basis.

\medskip\noindent\emph{Case A: $z_1z_2>0$.}
Then for any index $t$, the terms $z_1a_t$ and $z_2b_t$ have the same sign if $a_tb_t>0$, and opposite signs if $a_tb_t<0$.
\begin{itemize}
    \item  If Condition~($\star_1$) holds, then \((a_i b_i)(a_j b_j) \le 0\). Thus, at least one of \(a_i b_i\) and \(a_j b_j\) is non-negative.
    If $a_ib_i\ge 0$, then $z_1a_i$ and $z_2b_i$ have the same sign, so
    \[
    \|\mathbf{v}\|\ge |z_1a_i+z_2b_i| \ge |a_i| = \|\mathbf{a}\|.
    \]
   Else, $a_jb_j\ge 0$; using index $j$ gives
    \[
    \|\mathbf{v}\|\ge|z_1a_j+z_2b_j| \ge |b_j| = \|\mathbf{b}\|\ge\|\mathbf{a}\|.
    \]
    \item If Condition~($\star_2$) holds, then $(a_ib_i) (a_jb_j)>0$. 
    If $a_ib_i>0$, the same argument with index $i$ yields $\|\mathbf{v}\|\ge|z_1a_i+z_2b_i|\ge \|\mathbf{a}\|$.
    If $a_ib_i<0$, then both $a_ib_i$ and $a_jb_j$ are negative. By Condition~($\star_2$) there exists $k\neq i,j$ such that $a_ib_i a_kb_k\le0$ and $|a_k|+|b_k|\ge\|\mathbf{a}\|$. Because $a_ib_i<0$, we have $a_kb_k\ge0$, and therefore $z_1a_k$ and $z_2b_k$ share the same sign. Hence
    \[
    \|\mathbf{v}\|\ge |z_1a_k+z_2b_k| \ge |a_k|+|b_k| \ge \|\mathbf{a}\|.
    \]
\end{itemize}

\noindent\emph{Case B: $z_1z_2<0$.}
Here for any index $t$, if $a_tb_t\le0$, then $z_1a_t$ and $z_2b_t$ have the same sign; if $a_tb_t>0$, they have opposite signs.
\begin{itemize}
    \item If Condition~($\star_1$) holds, then $(a_ib_i) (a_jb_j)\le0$. Thus at least one of $a_ib_i$, $a_jb_j$ is non‑positive.
    If $a_ib_i\le0$, then $z_1a_j$ and $z_2b_j$ have the same sign and
      \[
   \|\mathbf{v}\|\ge |z_1a_i+z_2b_i| \ge |a_i| = \|\mathbf{a}\|.
    \]
   Else, $a_jb_j\le0$, and we obtain
     \[
   \|\mathbf{v}\|\ge |z_1a_j+z_2b_j| \ge |b_j| = \|\mathbf{b}\|\ge\|\mathbf{a}\|.
    \]
 
    \item If Condition~($\star_2$) holds, then $(a_ib_i) (a_jb_j)>0$.
    If $a_ib_i<0$ (both $a_ib_i$ and $a_jb_j$ are negative), the index $i$ immediately gives
    \[
    \|\mathbf{v}\|\ge|z_1a_i+z_2b_i| \ge |a_i| = \|\mathbf{a}\|.
    \]
    If $a_ib_i>0$, then $a_jb_j>0$ as well. Condition~($\star_2$) supplies an index $k\neq i,j$ with $a_ib_i a_kb_k\le0$ and $|a_k|+|b_k|\ge\|\mathbf{a}\|$. Since $a_ib_i>0$, we have $a_kb_k\le0$, so $z_1a_k$ and $z_2b_k$ share the same sign and
    \[
   \|\mathbf{v}\|\ge |z_1a_k+z_2b_k| \ge |a_k|+|b_k| \ge \|\mathbf{a}\|.
    \]
\end{itemize}

In every subcase we have exhibited a coordinate whose absolute value is at least $\|\mathbf{a}\|$. Consequently, $\|\mathbf{v}\|\ge\|\mathbf{a}\|$ for all nonzero $\mathbf{v}\in\mathcal{L}$, which proves $\lambda_1(\mathcal{L})=\|\mathbf{a}\|$.

\medskip\noindent\textbf{Argument for the length of $\mathbf{b}$}.
From the first part we already know $\|\mathbf{a}\|\le\min\{\|\mathbf{a}+\mathbf{b}\|,\|\mathbf{a}-\mathbf{b}\|\}$. If both $\|\mathbf{a}+\mathbf{b}\|\ge\|\mathbf{b}\|$ and $\|\mathbf{a}-\mathbf{b}\|\ge\|\mathbf{b}\|$, then
\[
\|\mathbf{a}\|\le\|\mathbf{b}\|\le\min\{\|\mathbf{a}+\mathbf{b}\|,\|\mathbf{a}-\mathbf{b}\|\},
\]
which is precisely the Lagrange reduction condition under the infinity norm. According to \textbf{Lemma \ref{lem:Lag}}, it follows that $\lambda_2(\mathcal{L})=\|\mathbf{b}\|$. 

Otherwise, at least one of \(\|\mathbf{a}+\mathbf{b}\|,\|\mathbf{a}-\mathbf{b}\|\) is strictly smaller than \(\|\mathbf{b}\|\). That is,
\[
\|\mathbf{a}\|\le\min\{\|\mathbf{a}+\mathbf{b}\|,\|\mathbf{a}-\mathbf{b}\|\}<\|\mathbf{b}\|.
\]
Recall that \(j\) is an index for which \(\|\mathbf{b}\|=|b_j|\).
\begin{itemize}
  \item If \(a_jb_j\ge0\), then \(|a_j+b_j| = |a_j|+|b_j| \ge |b_j|=\|\mathbf{b}\|\). Consequently,
  \[
  \|\mathbf{a}+\mathbf{b}\| \ge |a_j+b_j| \ge \|\mathbf{b}\|>\|\mathbf{a}-\mathbf{b}\| \ge   \|\mathbf{a}\| .
  \]

  \item If \(a_jb_j<0\), then \(|a_j-b_j| = |a_j|+|b_j| \ge \|\mathbf{b}\|\), giving   
  \[
  \|\mathbf{a}-\mathbf{b}\|\ge \|\mathbf{b}\|  > \|\mathbf{a}+\mathbf{b}\| \ge \|\mathbf{a}\|.
  \]
\end{itemize}

\end{proof}

\section{New Algorithm And Its Analysis}

 Based on the definition and properties of reduced bases for rank‑2 lattices under the infinity norm presented in Section \ref{sec:RedB}, we propose a reduction algorithm that avoids inner‑product computations. The core of this algorithm is a novel reduction approach that finds a shorter vector \(\mathbf{c} = \mathbf{b} - q\mathbf{a}\) by means of a newly designed quotient \(q\), which iteratively reduces the infinity norm of the basis vectors. 

For clarity, throughout this section, whenever an index $k$ satisfies $b_k\neq 0$, we write
\[
q_k=\left\lfloor \frac{a_k}{b_k}\right\rfloor_o,
\
r_k=a_k-q_k b_k.
\]
Thus, $q_s$ and $q_p$ denote the coordinate-wise quotients associated with the $s$-th and $p$-th coordinates, respectively. The indices $s,p,t$ are not fixed in advance; they are introduced in each case to denote coordinates at which the relevant infinity norm is attained, or coordinates whose signs are used to verify the reduction condition.

\begin{lemma}\label{lem:key}
Let $\mathbf{B}=[\mathbf{a}\  \mathbf{b}]$ be the lattice basis defined in
(\ref{eq:basis}) with
$\|\mathbf{a}\|=|a_i|, \|\mathbf{b}\|=|b_j|, |a_i|\ge |b_j|,$ where $i$ and $j$ may or may not be equal, and suppose that $\mathbf{B}$ is
not reduced. Define
$$
q_j=\left\lfloor\frac{a_j}{b_j}\right\rfloor_o,
\
q_j'=\operatorname{sgn}(a_jb_j)(|q_j|+1)=q_j+\operatorname{sgn}(a_jb_j),
$$
and let $\mathbf c=\mathbf a-q_j\mathbf b$, and $ \mathbf d=\mathbf a-q_j'\mathbf b$. Then $[\mathbf b\ \mathbf c], [\mathbf b\ \mathbf d]$ and $[\mathbf c \ \mathbf d]$ 
are valid lattice bases, and at least one of the following properties holds:

\begin{enumerate}
\item[(1)]
The vector $\mathbf c$ satisfies the length reduction condition
\begin{align}\label{eq:vc}
\|\mathbf c\|
<
\min
\left\{
\frac12\|\mathbf a\|,
\|\mathbf b\|
\right\}.
\end{align}

\item[(2)]
At least one of the bases $
[\mathbf b\ \mathbf c],\ 
[\mathbf b\ \mathbf d],\ 
[\mathbf c\ \mathbf d]
$
is reduced
\end{enumerate}
\end{lemma}

\begin{proof}
Let $
A=\|\mathbf a\|=|a_i|,\
B=\|\mathbf b\|=|b_j|,
$
so that $A>B$. Since $\mathbf B=[\mathbf a\ \mathbf b]$ is not reduced, applying Definition~\ref{def:Eucredn} to the ordered basis $[\mathbf b\ \mathbf a]$ gives $ a_i b_i a_j b_j>0.$

By replacing $\mathbf a$ (resp. $\mathbf b$) with $-\mathbf a$ (resp. $-\mathbf b$) if necessary, we may assume
$a_jb_j>0$ and $a_j>0, b_j>0$. This operation does not change the lattice or the reduction
status.  Consequently,
$$
q_j=\left\lfloor\frac{a_j}{b_j}\right\rfloor_o
=\left\lfloor\frac{a_j}{b_j}\right\rfloor
\geq0,  \
q_j'=q_j+1\ge 1.
$$
Since
$\mathbf c=\mathbf a-q_j\mathbf b,  \ \mathbf d=\mathbf a-(q_j+1)\mathbf b. $
Then
 $\mathbf c-\mathbf d=\mathbf b. $
Writing $a_j=q_jb_j+r, \ 0\leq r<b_j,$ we obtain
$c_j=r,\  d_j=r-b_j, $ and hence 
\begin{align}\label{eq:cd}
|c_j|+|d_j|=b_j=B, \  c_j\geq0,\ d_j<0. 
\end{align}

The basis validity follows from the unimodular transformations

$$
[\mathbf b\ \mathbf c]
=
[\mathbf a\ \mathbf b]
\begin{pmatrix}
0&1\\
1&-q_j
\end{pmatrix},
\
[\mathbf b\ \mathbf d]
=
[\mathbf a\ \mathbf b]
\begin{pmatrix}
0&1\\
1&-(q_j+1)
\end{pmatrix},
$$
and
$$
[\mathbf c\ \mathbf d]
=
[\mathbf a\ \mathbf b]
\begin{pmatrix}
1&1\\
-q_j&-(q_j+1)
\end{pmatrix},
$$
whose determinants are all $\pm1$.
It remains to prove the claimed reduction property.

\medskip
\noindent\textbf{Case 1.}
Suppose that the length of $\mathbf{c}$ satisfies \eqref{eq:vc}. Then the first branch is selected, and the desired conclusion follows immediately.

\medskip
\noindent\textbf{Case 2.}
Suppose that either $[\mathbf b\ \mathbf c]$ or $[\mathbf b\ \mathbf d]$ is reduced. Then the desired conclusion also follows immediately.

\medskip
\noindent\textbf{Case 3.}
It remains to consider the case where
\begin{align}\label{eq:cab}
\|\mathbf c\|
\geq
\min\left\{
\frac12A,B
\right\},
\end{align}
and neither $[\mathbf b\ \mathbf c]$ nor $[\mathbf b\ \mathbf d]$ is reduced. We prove that $[\mathbf c\ \mathbf d]$ must be reduced.

First, the $j$-th coordinate cannot attain $\|\mathbf c\|$.
Indeed, if $|c_j|=\|\mathbf c\|$, then $\|\mathbf c\|=r$. If $|a_j|<A$, then, since
$a_i b_i>0$ by the assumption,
$$
|c_i|=|a_i-q_jb_i|= A-q_j |b_i| \geq A-q_jB >|a_j|-q_jB=r=c_j,
$$
a contradiction. If $|a_j|=A$, then $q_j\geq1$. Hence
$$
A=|a_j|=|q_jb_j+c_j|=q_jb_j+r\ge b_j+r>2r\implies r<\frac A2;
$$
Meanwhile $r<b_j=B,$  which indicates
$$
\|\mathbf c\|=|c_j|= r<
\min\left\{\frac A2,B\right\},
$$
contradicting~(\ref{eq:cab}). Thus, if
$C=\|\mathbf c\|,$
we may choose an index $s\neq j$ such that
$
|c_s|=C.
$

Since $[\mathbf b\ \mathbf c]$ is not reduced and
$b_jc_j\ge 0$ by~(\ref{eq:cd}), the first condition of the reduction criterion fails,
and therefore 
\begin{align}\label{eq:s}
b_sc_s>0. 
\end{align}
First, we claim that 
\begin{align}\label{eq:s2}
c_sd_s\ge 0.
\end{align} 
Otherwise, $c_sd_s<0$ gives
$
B=|b_j|\ge|b_s|=|c_s-d_s|=|c_s|+|d_s|>C.
$
If $q_j=0$, then $\mathbf c=\mathbf a$ and hence $C=A$, yielding
$|b_s|>C=A>B$, a contradiction. If $q_j\geq1$, then, by~(\ref{eq:s}), we have
$$
A=|a_i|\ge|a_s|=|c_s+q_jb_s|=C+q_j|b_s|>2C.
$$
Thus $C<\min\{\frac12 A,B\}$, contradicting~(\ref{eq:cab}).

Let $D=\|\mathbf d\|,$ and choose an index $p$ such that $|d_p|=D.$  We now argue according to the value of $p$.

\medskip
\noindent\textit{Subcase 3.1:} $p=j$. Then the dominant coordinates of $\mathbf c$ and $\mathbf d$
are $s$ and $j$, respectively. Since $c_jd_j\le 0$ by (\ref{eq:cd}) and $c_sd_s\ge 0$ by (\ref{eq:s2}), we have 
$c_jd_jc_sd_s\le 0$ and thus, by $(\star_1)$, $[\mathbf c\ \mathbf d]$ is reduced.

\medskip
\noindent\textit{Subcase 3.2:} $p=s$. Then $\|\mathbf b\|=|b_j|$, $b_j>0$, $d_j<0$ and the non-reduction of
$[\mathbf b\ \mathbf d]$ imply
$b_sd_s<0.$ Together with (\ref{eq:s}), we have $c_sd_s<0$,  contradicting~(\ref{eq:s2}).
Therefore, $ p\neq s. $

\medskip
\noindent\textit{Subcase 3.3:} $p\neq j,s$. Then the dominant coordinates of $\mathbf c$ and $\mathbf d$
are $s$ and $p$, respectively. If $c_pd_p\le0$, then, by (\ref{eq:s2}), $c_sd_sc_pd_p\le 0$ and thus, by $(\star_1)$, $[\mathbf c\ \mathbf d]$ is reduced. If $c_pd_p>0$. Together with the condition $b_pd_p<0$ derived from the non-reduction of $[\mathbf b\ \mathbf d]$, we have 
\begin{align}\label{eq:bcdcase3}
b_pc_p<0\ \wedge |d_p|=|c_p-b_p|=|c_p|+|b_p|.
\end{align}
Since  $b_jc_jb_pc_p\le 0$, the non-reduction of $[\mathbf b\ \mathbf c]$ gives
$
|b_p|+|c_p|<\min\{B,C\}. 
$
Thus, the equation (\ref{eq:bcdcase3}) implies $$
D=\|\mathbf d\|=|d_p|<\min\{B,C\}\leq B.
$$
Using~(\ref{eq:cd}) again,
$$
|c_j|+|d_j|=B>D\ge\min\{C,D\},
$$
while $c_jd_jc_sd_s\ge 0$. Hence, by $(\star_2)$,  $[\mathbf c\ \mathbf d]$ is reduced.
 
Combining the above three cases, we complete the proof of the Lemma.
\end{proof}
 
\begin{algorithm}[ht]
\caption{\textbf{INFRed}$(\mathbf a,\mathbf b)$}
\label{alg:INFRed}
\begin{algorithmic}[1]

\REQUIRE A lattice basis $[\mathbf a\ \mathbf b]$ with
$\mathbf a,\mathbf b\in\mathbb Z^n$.
\ENSURE A reduced basis $[\mathbf a\ \mathbf b]$ under the $\ell_\infty$ norm.

\WHILE{$[\mathbf a\ \mathbf b]$ is not reduced}

    \IF{$\|\mathbf a\| < \|\mathbf b\|$}
        \STATE swap$(\mathbf a,\mathbf b)$
    \ENDIF

    \STATE Find indices $i,j$ such that
    $|a_i|=\|\mathbf a\|$ and
    $|b_j|=\|\mathbf b\|$.

    \STATE Compute
    $q\gets\left\lfloor a_j/b_j\right\rfloor_o$.

    \STATE $\mathbf c\gets\mathbf a-q\mathbf b$.

    \IF{$\|\mathbf c\|
    <\min\{\frac12\|\mathbf a\|,\|\mathbf b\|\}$}

        \STATE $[\mathbf a\ \mathbf b]\gets[\mathbf b\ \mathbf c]$

    \ELSE

        \IF{$[\mathbf b\ \mathbf c]$ is reduced}

            \RETURN $[\mathbf b\ \mathbf c]$

        \ELSE

            \STATE Compute
            $q'\gets\operatorname{sgn}(a_jb_j)(|q|+1)$.

            \STATE $\mathbf d\gets\mathbf a-q'\mathbf b$.

            \IF{$[\mathbf b\ \mathbf d]$ is reduced}

                \RETURN $[\mathbf b\ \mathbf d]$

            \ELSE

                \RETURN $[\mathbf c\ \mathbf d]$

            \ENDIF

        \ENDIF

    \ENDIF

\ENDWHILE

\RETURN $[\mathbf a\ \mathbf b]$

\end{algorithmic}
\end{algorithm}

Based on Lemma~\ref{lem:key}, we present an efficient lattice basis reduction
algorithm \textbf{INFRed} (Algorithm~\ref{alg:INFRed}). Its correctness and complexity
are established by the following theorem.

\begin{theorem}[Correctness and Complexity of \textbf{INFRed}]
\label{thm:INFRed}
Let $\mathbf B_0=[\mathbf a_0\ \mathbf b_0]$ be a basis of a
rank-two lattice in $\mathbb Z^n$, and let
\begin{align}\label{eq:T}
T=
\left\lceil
\log_2\max\{\|\mathbf a_0\|,\|\mathbf b_0\|\}
\right\rceil,
\end{align}
where $\|\cdot\|$ denotes the $\ell_\infty$-norm.
Then Algorithm~\ref{alg:INFRed} terminates and outputs a reduced
basis of the same lattice in
$O\bigl(nT^2\bigr)$
bit operations.
\end{theorem}

\begin{proof}
We first prove the correctness of Algorithm~\ref{alg:INFRed}.
Suppose that the current basis $[\mathbf a\ \mathbf b]$ is not
reduced. After possibly exchanging $\mathbf a$ and $\mathbf b$, we
may assume that $A=\|\mathbf a\|>\|\mathbf b\|=B.$
According to the algorithm,
$q_j=\left\lfloor\frac{a_j}{b_j}\right\rfloor_o$
and $\mathbf c=\mathbf a-q_j\mathbf b.$

If the length of $\mathbf c$ satisfies the equation (\ref{eq:vc}),
Algorithm~\ref{alg:INFRed} replaces
$[\mathbf a\ \mathbf b]$ by $[\mathbf b\ \mathbf c]$.
In particular,
$
\max\{\|\mathbf b\|,\|\mathbf c\|\}
=
B
<
A
=
\max\{\|\mathbf a\|,\|\mathbf b\|\}.
$
Thus the maximum norm of the two basis vectors strictly
decreases.

Otherwise, if $[\mathbf b\ \mathbf c]$ is reduced, the algorithm
terminates and returns this reduced basis. If
$[\mathbf b\ \mathbf c]$ is not reduced, the algorithm computes
$ q'_j=q_j+\operatorname{sgn}(a_jb_j) $ and $\mathbf d=\mathbf a-q'_j\mathbf b.$ By Lemma \ref{lem:key} at least one
of $[\mathbf b\ \mathbf d]$ and $[\mathbf c\ \mathbf d]$ is reduced, and all transformations performed by the
algorithm preserve the lattice. Therefore, whenever the algorithm terminates, it outputs a reduced
basis of the same lattice.

We next prove termination. For the
successive bases generated by the nonterminal iterations, let
$[\mathbf a_{i-1}\ \mathbf b_{i-1}]$
denote the input basis of the $i$-th iteration. Define
\[
t_0=
\left\lceil
\log_2\|\mathbf a_0\|
\right\rceil,
\
t_1=
\left\lceil
\log_2\|\mathbf b_0\|
\right\rceil.
\]
Whenever the algorithm does not terminate in the current iteration,
it is necessarily in the case
\[
\|\mathbf c_i\|
<
\min\left\{
\frac{\|\mathbf a_{i-1}\|}{2},
\|\mathbf b_{i-1}\|
\right\},
\]
and consequently
$
[\mathbf a_i\ \mathbf b_i]
=
[\mathbf b_{i-1}\ \mathbf c_i].
$
Therefore,
\begin{align}
\label{eq:norm-recurrence-1}
\|\mathbf a_i\| =
\|\mathbf b_{i-1}\|, \  
\|\mathbf b_i\|
=
\|\mathbf c_i\|
<
\min\left\{
\frac{\|\mathbf a_{i-1}\|}{2},
\|\mathbf b_{i-1}\|
\right\}.
\end{align}
It follows from~\eqref{eq:norm-recurrence-1} that
\begin{align}
\label{eq:bit-recurrence}
t_i=
\left\lceil
\log_2\|\mathbf a_i\|
\right\rceil
=
\left\lceil
\log_2\|\mathbf b_{i-1}\|
\right\rceil,\ 
t_{i+1}
\leq
t_i
\ \text{and}\ 
t_{i+1}
\leq
t_{i-1}-1.
\end{align}
Thus, the bit length decreases by at least one every two
consecutive nonterminal iterations. In particular, the number of
nonterminal iterations is $O(T)$, and Algorithm~\ref{alg:INFRed}
terminates.

Finally, we argue the claimed bit complexity. Consider the $i$-th nonterminal iteration and write
$
A_{i-1}=\|\mathbf a_{i-1}\|,
\ 
B_{i-1}=\|\mathbf b_{i-1}\|.
$
Since $|a_j|\leq A_{i-1}$ and $|b_j|=B_{i-1}$, we have
\[
|q_j|
=
\left|
\left\lfloor
\frac{a_j}{b_j}
\right\rfloor_o
\right|
\leq
\frac{A_{i-1}}{B_{i-1}}.
\]
Hence, if $
\Delta_i=t_{i-1}-t_i+1,
$
then $q_j$ has at most $O(\Delta_i)$ bits. The same bound holds for
$q'_j$, since $q'_j=q_j+\operatorname{sgn}(a_jb_j).$
The computation of the quotient $q_j$ and the remainder can be
performed by integer division whose bit complexity is
\[
O(t_i\Delta_i)
\]
under the standard quadratic-time division model, since the divisor
has $t_i$ bits and the quotient has $O(\Delta_i)$ bits.

For the vector update
$
\mathbf c_i
=
\mathbf a_{i-1}-q_j\mathbf b_{i-1},
$
each coordinate requires a multiplication of a
$t_i$-bit integer by an $O(\Delta_i)$-bit integer, followed by an
addition. Under quadratic-time integer multiplication, the cost per
coordinate is
$
O(t_i\Delta_i).
$
Thus, computing the entire $n$-dimensional vector $\mathbf c_i$
requires
$
O(nt_i\Delta_i)
$
bit operations. The same bound applies to the computation of
$\mathbf d_i$ when it is required.

Consequently, the total bit complexity of the $i$-th iteration is
$
O(nt_i\Delta_i).
$
Since $t_i\leq T$, we obtain the total complexity of the algorithm is
\begin{align*}
&\sum_i nt_i\Delta_i\leq
nT\sum_i\Delta_i =
nT
\sum_i
(t_{i-1}-t_i+1)\\
= &nT(\sum_i(t_{i-1}-t_i)+\sum_i1)=nT(t_0-t_\ell+\sum_i1)=nT\cdot O(T)=(nT^2)
\end{align*}
\end{proof}

\section{Practical Performance Evaluation}

In this section, we implement the proposed  lattice basis reduction algorithm \textbf{INFRed} (Algorithm \ref{alg:INFRed}) and evaluate its practical performance. 

\subsection{Experimental Methodology}

To comprehensively evaluate the computational performance of the proposed
algorithm, \textbf{INFRed}, we consider two experimental settings. The
first evaluates the computation of the shortest basis under the
$\ell_\infty$ norm, while the second investigates whether \textbf{INFRed}
can serve as an effective preprocessing procedure for Euclidean lattice
reduction under the $\ell_2$ norm.

\vspace{0.15cm}\noindent\textbf{Comparison under the $\ell_\infty$ norm.}
For a rank-two lattice $\mathcal{L}$, let $\lambda_1$ and $\lambda_2$
denote the lengths of the first and second successive minima under the
$\ell_\infty$ norm, respectively. We compare two approaches for computing
a shortest basis corresponding to $(\lambda_1,\lambda_2)$.

The first approach uses the proposed \textbf{INFRed} algorithm to obtain
a shortest vector of length $\lambda_1$. Since \textbf{INFRed} directly
provides $\lambda_1$ but does not necessarily produce a second basis
vector of length $\lambda_2$, we subsequently perform one additional
Lagrange--Gauss reduction based on a binary search to obtain $\lambda_2$.
The second approach directly applies the Lagrange--Gauss algorithm with
binary search to compute both $\lambda_1$ and $\lambda_2$. Thus, the
comparison measures whether the proposed reduction strategy can avoid
the repeated binary-search procedures required by the conventional
approach when computing the shortest basis.

\vspace{0.15cm}\noindent\textbf{Comparison under the $\ell_2$ norm.}
We further investigate whether \textbf{INFRed}, although designed for the
$\ell_\infty$ norm, can be used to accelerate shortest-basis computation
under the Euclidean norm. To this end, we compare two approaches. The
first applies \textbf{INFRed} to the input basis as a preprocessing step
and then applies the classical Lagrange--Gauss algorithm to the resulting
shorter basis to obtain the final shortest basis under the $\ell_2$ norm.
The second applies the Lagrange--Gauss algorithm directly to the original
basis without preprocessing. 
This experiment is intended to evaluate whether the rapid reduction of
basis-vector magnitudes achieved by \textbf{INFRed} can compensate for
the additional preprocessing cost and thereby reduce the overall cost
of subsequent Euclidean lattice reduction.

\subsection{Experimental Environment and Settings}

All experiments were conducted on Ubuntu 22.04 running in a VMware virtualized environment with 4 GB RAM. The proposed algorithm and all baseline methods were implemented in C and compiled using GCC with optimization enabled. 

The efficiency of lattice reduction algorithms is strongly affected by the linear dependence between the input basis vectors. We therefore generate test lattices with different dimensions and degrees of linear
dependence to evaluate the algorithms under a range of input conditions.

Consider a rank-two lattice basis
$
\mathbf{B}=[\mathbf{a},\mathbf{b}],
$
where $
\mathbf{a}=(a_1,a_2,\ldots,a_n)^T,\
\mathbf{b}=(b_1,b_2,\ldots,b_n)^T.
$
 We measure the linear dependence
between $\mathbf{a}$ and $\mathbf{b}$ by
\[
\Delta(\mathbf{a},\mathbf{b})
=
\max_{\substack{i,j\in\{1,\ldots,n\}\\b_i b_j\neq 0}}
\left|
\frac{a_i}{b_i}-\frac{a_j}{b_j}
\right|.
\]
A smaller value of $\Delta(\mathbf{a},\mathbf{b})$ indicates that the
coordinate-wise ratios $a_i/b_i$ are more concentrated, and hence that
the two basis vectors exhibit stronger linear dependence. To generate test instances with controlled linear dependence, we use
the random continued-fraction expansions of the coordinate ratios $a_i/b_i$.
In particular, for two coordinates $i,j\in\{1,\ldots,n\}$, we control
the number of common leading terms in the continued-fraction expansions
of $a_i/b_i$ and $a_j/b_j$. We denote by $\kappa$ the proportion of
identical terms, counted from the beginning of the continued-fraction
expansions, relative to the total number of expansion terms. The
parameters $\delta$ and $\kappa$ are therefore used to characterize the
degree of linear dependence: we generate instances with
$\Delta\approx 10^{-\delta},$
where a larger $\delta$ corresponds to a smaller $\Delta$ and hence a
stronger linear dependence. Increasing $\kappa$ likewise increases the
length of the common prefix of the continued-fraction expansions and
therefore produces more strongly correlated basis vectors.

For each combination of the lattice dimension $n$ and the linear
dependence parameters, we generate 10 random lattice instances. The
maximum bit length of the coordinates of $\mathbf{a}$ and $\mathbf{b}$
is fixed at 5000 bits. The reported execution time is the average
running time over the 10 instances.

\subsection{Experimental Results and Analysis}

\paragraph{Results under the $\ell_\infty$ norm.} Table~\ref{tab:alg2_vs_lagrange_infty} compares the running times of \textbf{INFRed} and the Lagrange--Gauss algorithm for computing
$\lambda_1$ and $\lambda_2$ under the $\ell_\infty$ norm. In particular, the column ``INFRed'' represents the running time of \textbf{INFRed} for obtaining $\lambda_1$, followed by one additional binary-search-based reduction for obtaining $\lambda_2$, whereas the ``L-G'' column represents the running time of the Lagrange--Gauss algorithm for computing both successive minima.
Across all tested dimensions and linear-dependence levels, \textbf{INFRed} is
consistently faster than the Lagrange--Gauss approach. The observed
speedup ranges from approximately $4.44\times$ to $7.32\times$.

For a fixed dimension, the running time of both methods generally
increases as the dimension grows. For example, when
$\delta=375$, the running time of \textbf{INFRed} increases from
$0.0045$~s for $n=100$ to $0.0249$~s for $n=500$, while the corresponding
running time of Lagrange--Gauss increases from $0.0303$~s to
$0.1806$~s. This behavior is consistent with the linear dependence of
the per-iteration vector operations on the ambient dimension.

The results also show that the speedup remains substantial as the degree
of linear dependence varies. Although the exact speedup fluctuates with
$\delta$, \textbf{INFRed} maintains a speed advantage over the
Lagrange--Gauss approach for all tested parameter settings.

\paragraph{Results under the $\ell_2$ norm.}
Table~\ref{tab:alg2_vs_lagrange_l2} evaluates the use of \textbf{INFRed}
as a preprocessing step for Euclidean lattice reduction. The column
``INFRed + L-G'' denotes the total running time of first applying
\textbf{INFRed} and then applying Lagrange--Gauss to the resulting basis.
For comparison, the ``L-G'' column reports the running time of applying
Lagrange--Gauss directly to the original basis.

The results demonstrate a substantial reduction in the overall running
time when \textbf{INFRed} is used as a preprocessing procedure. The
observed speedup ranges from approximately $21.41\times$ to
$41.93\times$ across the tested instances. For example, when
$n=500$ and $\delta=375$, the direct Lagrange--Gauss approach requires
$1.2784$~s, whereas the combination of \textbf{INFRed} and
Lagrange--Gauss requires only $0.0366$~s, corresponding to a speedup of
approximately $34.93\times$.

This improvement can be attributed to two complementary effects.
First, \textbf{INFRed} performs reduction using only coordinate-wise
operations under the $\ell_\infty$ norm and therefore avoids the
large-integer inner-product computations required by the Euclidean
Lagrange--Gauss algorithm. Second, \textbf{INFRed} rapidly decreases the
magnitudes of the basis vectors before the subsequent Euclidean
reduction. Consequently, the Lagrange--Gauss algorithm operates on a
substantially smaller basis, reducing the computational cost of the
subsequent $\ell_2$-norm reduction.

Overall, the experiments indicate that \textbf{INFRed} provides a
consistent computational advantage for $\ell_\infty$-norm shortest-basis
computation and can also serve as an effective preprocessing technique
for Euclidean lattice reduction. The latter result suggests that the
proposed $\ell_\infty$ reduction framework may be useful beyond the
direct computation of $\ell_\infty$-shortest vectors.

\begin{table}[htp]
	\centering

	\caption{Running-time comparison for computing $\lambda_1$ and
$\lambda_2$ under the $\ell_\infty$ norm. \textbf{INFRed} first computes
$\lambda_1$ and then uses one additional binary-search-based reduction to
compute $\lambda_2$, whereas Lagrange--Gauss (\textbf{L\_G}) computes both
successive minima using binary search. (Unit: seconds.)}	
	\label{tab:alg2_vs_lagrange_infty}
	\setlength{\tabcolsep}{5pt}
	\resizebox{\textwidth}{!}{
	\begin{tabular}{c c c c r r r} 
		\toprule
		\multicolumn{4}{c}{Lattice Parameters} & \multicolumn{3}{c}{Execution Time (s)} \\ 
		\cmidrule(lr){1-4} \cmidrule(lr){5-7} 
		\shortstack{$\delta$} & \shortstack{Dimension ($n$)} & \shortstack{bitsize ($T$)} & $\kappa$ 
		& \textbf{INFRed} & \multicolumn{1}{c}{\textbf{L\_G}} & \shortstack{\textbf{L\_G/INFRed}} \\
		\midrule
		375 & 100  & 5000  & 0.125  & 0.0045   & 0.0303  & 6.7333 \\
		750 & 100  & 5000  & 0.25  & 0.0105   & 0.0586  & 5.5810 \\
		1500 & 100  & 5000  & 0.5  & 0.0205   & 0.1019  & 4.9707 \\
		2250 & 100  & 5000  & 0.75  & 0.0266   & 0.1382  & 5.1955 \\
		2625 & 100  & 5000  & 0.875   & 0.0316   & 0.1404  & 4.4430 \\
		\addlinespace
		375& 300  & 5000  & 0.125   & 0.0171   & 0.1032  & 6.0351 \\
		750& 300  & 5000  & 0.25   & 0.0267   & 0.1923  & 7.2022 \\
		1500& 300  & 5000  & 0.5   & 0.0467   & 0.3263  & 6.9872 \\
		2250& 300  & 5000  & 0.75   & 0.0681   & 0.4078  & 5.9882 \\
		2625& 300  & 5000  & 0.875   & 0.0789   & 0.4327  & 5.4842 \\
		\addlinespace
		375& 500  & 5000  & 0.125   & 0.0249   & 0.1806  & 7.2530 \\
		750& 500  & 5000  & 0.25   & 0.0433   & 0.3170  & 7.3210 \\
		1500& 500  & 5000  & 0.5   & 0.0877   & 0.5341  & 6.0901 \\
		2250& 500  & 5000  & 0.75   & 0.1234   & 0.6762  & 5.4797 \\
		2625& 500  & 5000  & 0.875  & 0.1315   & 0.7089  & 5.3909 \\
		\bottomrule
	\end{tabular}
	}
\end{table}

\begin{table}[htp]
\centering
\caption{Running-time comparison for computing the shortest basis under
the $\ell_2$ norm. ``INFRed + L-G'' first applies \textbf{INFRed} as a
preprocessing step and then applies Lagrange--Gauss (\textbf{L\_G}) to
obtain the final $\ell_2$-shortest basis, while ``L-G'' applies
Lagrange--Gauss directly to the original basis. (Unit: seconds.)}
	\label{tab:alg2_vs_lagrange_l2}
\small
\resizebox{\textwidth}{!}{
\setlength{\tabcolsep}{5pt}
\begin{tabular}{c c c c r r r r} 
\toprule
\multicolumn{4}{c}{Lattice Parameters} & \multicolumn{4}{c}{Execution Time (s)} \\ 
\cmidrule(lr){1-4} \cmidrule(lr){5-8} 
\shortstack{$\delta$} & \shortstack{Dimension ($n$)} & \shortstack{bitsize ($T$)} & $\kappa$ 
& \textbf{INFRed} & \multicolumn{1}{c}{\textbf{L\_G}} & \shortstack{\textbf{INFRed} + \\ \textbf{L\_G}} & \shortstack{\textbf{L\_G} / \\ \textbf{(INFRed+L\_G)}}\\
\midrule
375& 100  & 5000  & 0.125  & 0.0049   & 0.2355  & 0.0066  & 35.6818\\
750& 100  & 5000  & 0.25   & 0.0086   & 0.4069  & 0.0104  & 39.1250\\
1500& 100  & 5000  & 0.125  & 0.0170   & 0.6542  & 0.0181  & 36.1436\\
2250& 100  & 5000  & 0.125  & 0.0262   & 0.7176  & 0.0265  & 27.0792\\
2625& 100  & 5000  & 0.125  & 0.0292   & 0.6768  & 0.0295  & 22.9424\\
\addlinespace
375& 300  & 5000  & 0.40   & 0.0157   & 0.7801  & 0.0222  & 35.1396\\
750& 300  & 5000  & 0.40   & 0.0272   & 1.3503  & 0.0322  & 41.9348\\
1500& 300  & 5000  & 0.40   & 0.0551   & 1.9136  & 0.0588  & 32.5442\\
2250& 300  & 5000  & 0.40   & 0.0745   & 2.1077  & 0.0758  & 27.8061\\
2625& 300  & 5000  & 0.60   & 0.0787   & 2.0300  & 0.0791  & 25.6637\\
\addlinespace
375& 500  & 5000  & 0.75   & 0.0234   & 1.2784  & 0.0366  & 34.9290\\
750& 500  & 5000  & 0.75   & 0.0488   & 2.2421  & 0.0581  & 38.5904\\
1500& 500  & 5000  & 0.75   & 0.0941   & 3.3891  & 0.0997  & 33.9930\\
2250& 500  & 5000  & 0.75   & 0.1241   & 3.4339  & 0.1278  & 26.8693\\
2625& 500  & 5000  & 0.875  & 0.1555   & 3.3464  & 0.1563  & 21.4101\\
\bottomrule
\end{tabular}
}
\end{table}

\section{Conclusion}

This paper presented a new lattice reduction framework for rank-2 lattices
in $\mathbb{R}^n$ under the $\ell_\infty$ norm. By exploiting the
coordinate-wise structural properties of lattice basis vectors, the
proposed approach avoids expensive inner-product computations and
achieves more efficient lattice reduction compared with existing
approaches. Both theoretical analysis and experimental evaluations
demonstrate its favorable computational performance, particularly for
highly correlated high-dimensional lattice bases. The proposed framework
provides a useful foundation and new insights for developing efficient
lattice basis reduction algorithms for higher-rank lattices.


\bibliographystyle{IEEEtran}
\bibliography{ref}

\newpage

\end{document}